\documentclass{llncs}
\usepackage{amsmath,amssymb,amsfonts}
\usepackage{algorithm}
\usepackage[noend]{algpseudocode}
\usepackage{adjustbox}
\usepackage[graphicx]{realboxes}
\usepackage{textcomp}
\usepackage{xcolor}
\usepackage{rotating}
\usepackage{pdflscape}
\usepackage{threeparttable}
\usepackage{hyperref}
\usepackage{booktabs}
\usepackage{pgfplots}
\usepackage{mathtools}
\usepackage{geometry}
\begin{document}
\title{PrivaMatch: A Privacy-Preserving\\ DNA Matching Scheme for Forensic Investigation}

\author{Sankha Das$^\dagger$}
\institute{BITS Pilani, India}



\maketitle


\begin{abstract}
DNA fingerprinting and matching for
identifying suspects has been a common practice in
criminal investigation. Such proceedings involve multiple parties such as investigating agencies, suspects and forensic labs.
A major challenge in such settings is to carry out the matching process
between the suspects’ DNA samples and the samples
obtained from the crime scene without compromising
the privacy of the suspects’ DNA profiles. Additionally,
it is necessary that sensitive details pertaining to the
investigation such as the identities of the suspects and
evidence obtained from the crime scene must be kept
private to the investigating agency. We present a novel DNA matching scheme,
termed as PrivaMatch, which addresses multiple concerns
about privacy of the suspects’ DNA profiles and the crime
scene evidence. In the proposed scheme, the investigating agencies oblivious transfer and zero-knowledge proofs
to privately obtain the DNA profiles of the suspects from the forensic lab's database. In addition, we present a 
clever data obfuscation technique using homomorphic encryption and modular arithmetic for the investigating agency to privately obtain the DNA profile of the crime scene's sample, keeping the profile oblivious from the forensic lab.
The DNA profile of the crime
scene sample is operated on using a homomorphic
cryptosystem such that neither of the parties (e.g., the
investigation agency, forensic labs,
DNA database owners) learns about the private data
of the other parties. The proposed scheme is analysed
formally and the practicality of its security strengths
is verified using simulations under standard assumptions.

\vspace{2mm}
\textbf{Keywords:} DNA Matching, Multiparty Computation, Homomorphic Encryption, Security, Privacy.

\end{abstract}



\section{Introduction}
\label{sec:intro}
Deoxyribonucleic Acid (DNA) \cite{Alberts2002} is the biochemical encoding of genetic material that is present in every cell of humans. The DNA of a person forms a basis for the unique identity of the person; the probability that the entire DNA sequence of two individuals in the world is identical is extremely low \cite{Koehler2001}. This uniqueness serves as the underlying principle for tasks such as paternity testing, identity establishment and genealogical database construction. DNA matching has been extensively studied during the past four decades for finding applications in forensic research, including criminal investigation. Typically, DNA samples are extracted from cells found in crime scene evidence such as the victim's body or the perpetrator's belongings left at the scene such as caps, shoes, rings, etc. The genetic material extracted from the cells needs to be purified in order to extract sufficient quantity of pure DNA, which can be used for the subsequent matching process(es) \cite{Chong2021}. Further, a DNA extraction step is also carried out to collect DNA samples of a group of suspects. These DNA samples are then compared with the sample obtained from the crime scene to check if there is a match, and if so, it indicates that the particular suspect is a likely perpetrator of the crime. The extracted DNA sample is purified using chemical reagents and special restriction enzymes are used to extract those segments of the DNA sequence which are unique for each individual \cite{Wang2011}. These segments are then subjected to an electrophoresis process followed by an X-ray exposure step which yields a pattern of DNA bands. The pattern of DNA bands is representative of the nucleotide sequence and represents the DNA profile unique to the individual. Newer methods such as STR analysis \cite{Kashyap2004} yield a DNA profile where fragments of different sizes in the extracted DNA produce signals of different amplitudes during the electrophoresis. For the purpose of this work, we represent the resulting DNA profile as a matrix composed of fragment sizes and the amplitude of the signal produced by the respective fragment. This profile can be represented in the below form.

\begin{equation}
    \mathbf{d} = \begin{bmatrix}
                    \eta_1 & \sigma_1 \\
                    \eta_2 & \sigma_2 \\
                    \vdots & \vdots \\
                    \eta_t & \sigma_t
                \end{bmatrix},
    \label{eq:1}
\end{equation}
where $\eta_i$ represents the fragment size and $\sigma_i$ represents the amplitude of the signal produced by the $i^{th}$ fragment or ``slot". For the purpose of this work, we will assume that the elements of a DNA profile $\mathbf{d}$ are positive integers, each of which is a member of a multiplicative field $\mathbb{Z}^*_p$ of order $p$ for a large prime number $p$. DNA matching is a crucial step in narrowing down the list of suspects and also in exonerating the wrongly accused. In the past, DNA matching has been used to conclude many criminal investigations involving murder, abduction, sexual assault, robbery and so on. The fact that the DNA of a person is a representation of the unique identity of the person combined with the finding that the probability of two persons having the same DNA profile is extremely rare, makes DNA evidence a very strong standard in criminal investigation proceedings and even in the court of law \cite{Butler2005}.

With the establishment of large DNA databases across the world \cite{Roewer2013} for scientific and forensic research, there arises a question of privacy of the individuals whose DNA profiles have been stored in such databases. While previous works \cite{Bruekers2008}\cite{Wang2017} have addressed the question of secure and privacy-preserving DNA matching for issues such as paternity testing and genealogical ancestry determination, secure DNA matching has many more facets to itself from the standpoint of criminal investigation. The latter problem is more complex due to several ethical, legal and technical issues. Firstly, DNA matching conducted against samples from such databases exposes the DNA profiles of the individuals to third parties such as investigation agencies and courts of law. Past works show that DNA samples can be fabricated in labs given the DNA profile of the individual \cite{Frumkin2010}, which makes the individual prone to being falsely framed in a crime by implanting the artificial DNA sample at the crime scene. To limit the possibility of such scenarios, it is necessary that comparisons on DNA profiles be carried out on the obfuscated or encrypted profiles rather than publicly available DNA profiles. Not only does this measure preserve the privacy and integrity of the DNA profile of the respective individual, but also reduces the occurrence of foul play in investigation and judicial procedures. Secondly, matching the DNA sample obtained from the crime scene against certain chosen suspects raises an alarm for the perpetrator of the crime, which might incite them to interfere with the investigation process. Even if the perpetrators themselves do not learn about the investigation proceedings, they can be apprised of the same by parties that are part of the investigation process (e.g. the forensic lab). Therefore, it is necessary that the list of suspects remains private to the investigation agency only. Furthermore, the DNA sample obtained from the crime scene has to be again encrypted under the same system as the profiles in the DNA database so as to allow comparison over encrypted samples. The encryption of this sample must also be done in a way that does not reveal the actual DNA profile of the sample to parties such as the forensic labs. Due to the presence of such additional concerns, the problem of secure DNA matching for criminal investigation is much more intricate than other issues.\vspace{1 mm}

\textbf{\textit{Our Contributions.}} We present a private DNA matching scheme, termed as \textbf{PrivaMatch}, which addresses the privacy and security concerns in DNA matching for forensic investigation. We consider a scenario in which a criminal investigation is in process. The entire matching process involves carrying out computations over these encrypted DNA samples using secure multi-party computation primitives \cite{Lindell2020}\cite{Cramer2015}. The investigation is carried out by an investigation agency $\mathcal{I}$ which is a semi-honest party \cite{Lindell2010}. $\mathcal{I}$ is entrusted with collecting evidence from the crime scene, pull up a list of suspects and check for a match between the DNA profiles of the suspects with a DNA sample that is collected from the evidence. The second party in the system is a forensic lab $\mathcal{L}$ which is a semi-honest party and has access to an encrypted database of DNA profiles of the citizens of the country. The privacy of individuals is preserved in the scheme by storing the DNA profile in a database encrypted under a homomorphic cryptosystem \cite{Yi2014}, the key for which is owned by $\mathcal{L}$. In the implementation of the proposed scheme, the database is encrypted under the BGV cryptosystem \cite{Brakerski2014}, which is a fully homomorphic encryption (FHE) cryptosystem that allows arbitrary number of modular additions and multiplications over the encrypted domain. The proposed scheme consists of three phases. The first phase involves obtaining the encrypted DNA samples of the suspects from the DNA profile database such that the owner of the database does not learn the identities of the specific suspects that the investigation agency is interested in. We use an oblivious transfer scheme \cite{Rabin2005}\cite{Even1985} involving asymmetric-key encryption and zero-knowledge proofs \cite{Goldreich1994} to obtain the DNA profiles of the suspects whom the investigation agency is suspecting. As a result, the owner of the DNA database does not learn which specific suspects are being targeted and the investigating agency learns the encrypted DNA profiles of only the suspects. The second phase uses an algorithm to privately obtain an encrypted profile of the DNA sample obtained from the crime scene using homomorphic encryption. The scheme does not want the owner of the DNA profile database (the Forensic Lab) to learn about the crime scene DNA sample, lest it should carry out a local DNA matching process against the entire database and alert the perpetrator(s) in case a match is found. The third and final phase calculates similarities between the crime scene DNA profile and each of the suspects' DNA profiles in the encrypted domain, while maintaining the privacy of the suspects' DNA profiles. This is done by calculating a distance-metric between the each of the suspects' DNA profiles and the crime scene DNA profile using arithmetic operations over the encrypted domain. A match is found by decrypting these distance metrics for each of the suspects' DNA profiles and comparing them against a public threshold decided by the investigation agency. All the suspects whose DNA profile's distance from the crime scene DNA profile falls within the threshold, are declared to be a match for the DNA sample obtained from the crime scene. The Forensic Lab does not learn about these specific suspects from the decrypted distance-metrics and is therefore unable to interfere with the investigation process. 

The remainder of the paper is organized as follows. Section \ref{sec:rel_work} discusses about the existing literature on secure DNA matching and past work done in this particular application of secure multiparty computation. Section \ref{sec:scheme} presents the proposed scheme, \textit{PrivaMatch}, explaining different stages of the scheme along with the detailed algorithms used in each phase. Section \ref{sec:sec_anal} provides security proofs of the proposed scheme followed by a performance summary in Section \ref{sec:perf_anal}. We conclude the paper in Section \ref{sec:concl}.

\begin{table}[ht]
    \centering
    \caption{List of Symbols and their Meanings}
    \vspace{1mm}
    \begin{tabular}{|c|p{2in}|}
        \hline
        \multicolumn{1}{|c|}{\textbf{Notation}}                                          & \multicolumn{1}{c|}{\textbf{Meaning}}\\ \hline
        $\mathcal{K}$         & Key space \\ \hline
        $\mathcal{M}$         & Message space \\ \hline
        $\mathcal{M'}$        & Ciphertext space \\ \hline
        $\mathcal{E}(m, k)$   & Encryption algorithm $\mathcal{E}: \mathcal{M} \times \mathcal{K} \to \mathcal{M'}$ 
                                where $m \in \mathcal{M}, k \in \mathcal{K} $ \\ \hline
        $\mathcal{D}(m', k)$  & Decryption Algorithm $\mathcal{D}: \mathcal{M'} \times \mathcal{K} \to \mathcal{M} $ 
                                where $m' \in \mathcal{M'}, k \in \mathcal{K}$ \\ \hline
        $\mathcal{C} = (\mathcal{E}, \mathcal{D})$    & Cryptosystem $\mathcal{C}$ with Encryption algorithm $\mathcal{E}$ and Decryption algorithm                                                            $\mathcal{D}$ \\ \hline
        $\mathbb{Z}^*_p$      & Multiplicative prime field of order $p$ \\ \hline
        $\mathbb{I}^j_i$      & Set of positive integers $\{i, i+1, \cdots, j\}$ \\ \hline
        $(S)_i$               & $i^{\text{th}}$ element of set/vector $S$ \\ \hline
        $S_{i..j}$            & $\{s_t: s_t \in S$ $\forall t \in \mathbb{I}^j_i\}$ \\ \hline
        $[\mathbf{M}]_{i,j}$    & Element in the $i^{\text{th}}$ row and $j^{\text{th}}$ column in matrix $\mathbf{M}$ \\ \hline
        $r \overset{\$}{\leftarrow} S$ & $r$ is sampled randomly from set $S$ \\ \hline
        $\mathcal{I}$         & Investigating Agency \\ \hline
        $\mathcal{L}$         & Forensic Lab \\ \hline
        $e^H$                 & Encryption of $e$ under homomorphic cryptosystem $\mathcal{C}_H = (\mathcal{E}_H, \mathcal{D}_H)$\\ \hline
    \end{tabular}
    \label{tab:symbols}
\end{table}

\section{Related Work}\label{sec:rel_work}
Over the past four decades, significant work has been done in the domain of secure MPC and its use in pattern matching, keyword search, and private text processing. In \cite{Tronocoso-Pastiroza2007}, the authors have described a robust privacy-preserving string search protocol for verifying the presence of a certain private genomic sequence in another private DNA sequence. Katz \& Malka in \cite{Katz2010} have proposed a secure keyword search protocol, where two parties $P_1$ and $P_2$ holding private inputs ``search database'' and ``keyword'', respectively, using which $P_2$ learns positions where the keyword can be found in the database without $P_1$ learning the keyword. As illustrated by the authors, the protocol has applications in pattern matching, including a use case for finding short-tandem repeats (STRs) in DNA sequences. Yasuda \textit{et al.} in \cite{Yasuda2014} describe another protocol for private wildcard pattern matching applied to searching DNA sequences in genomic data using Somewhat Homomorphic Encryption. In tackling a similar problem statement, De Cristofaro \textit{et al.} in \cite{Cristofaro2013} propose a cryptographic primitive called Size and Position-Hiding Private Substring Matching to privately check the presence of a genomic pattern in a patient's DNA sequence. The use of a distance-metric to calculate the similarities between the crime scene DNA profile and the DNA profiles of the suspects used in our work, is inspired by the work done by Drozdowski \textit{et al.} in \cite{Drodowski2019}. Here, the authors tackle the problem of private biometric face identification using homomorphic encryption to protect the privacy of the individuals' profiles and calculate similarities between the individuals' profiles and a new test profile using a distance metric calculated over the encrypted domain.

\section{PrivaMatch: The Proposed Scheme}\label{sec:scheme}
We now present the proposed scheme, \textit{PrivaMatch}, to find a DNA match in a DNA profile database $\mathbf{D}$ for the DNA profile $\mathbf{d_s}$ of the crime-scene sample. The database of DNA profiles $\mathbf{D}$ is encrypted under the BGV cryptosystem, $\mathcal{C}_H = (\mathcal{E}_H, \mathcal{D}_H)$, whose key $k_H$ is owned by $\mathcal{L}$. We assume that all communications take place over a secure channel. We now describe the steps and underlying protocols using which semi-honest parties $\mathcal{I}$ and $\mathcal{L}$ communicate with each other to find a match in $\mathbf{D}$ for $\mathbf{d_s}$. 
        
\subsection{Getting DNA Profiles of Suspects}
$\mathcal{I}$ suspects a group of $m$ people $S$ and obtains the DNA profiles of these suspects from $\mathcal{L}$ using an oblivious transfer scheme described in \cite{Even1985}. As a result of this scheme, $\mathcal{I}$ learns the DNA profiles of $s_i \in S$ encrypted under $\mathcal{C}_H$. Again, $\mathcal{L}$ does not learn which suspects $\mathcal{I}$ is interested in. To carry out this step, protocol $\mathbf{\Pi}_{\textsc{getprofiles}}$ is executed with $P_0 = \mathcal{I}$ and $P_1 = \mathcal{L}$. We note that in this protocol, an asymmetric-key cryptosystem $\mathcal{C} = (\mathcal{E}, \mathcal{D})$ is used which is different than the homomorphic cryptosystem $\mathcal{C}_H$.

        \begin{algorithm}[H]
            \caption{Get Suspect DNA Profiles $\mathbf{\Pi}_{\textsc{getprofiles}}$($P_0, P_1$)}\label{alg:4}
            \textbf{Input:} $P_0$ holds a set of suspects $S = \{s_1, s_2,\cdots, s_m\}$ and a set of $n$ public keys
                            $K = \{k_1, k_2,\cdots, k_n\}$ under an asymmetric key cryptosystem $\mathcal{C} = (\mathcal{E}, \mathcal{D})$.
                            $P_0$ holds private keys corresponding to $m$ of these public keys.
                            $P_1$ owns a database of $n$ DNA profiles encrypted under an FHE cryptosystem 
                            $\mathcal{C}_H = (\mathcal{E}_H, \mathcal{D}_H)$.
    
            \textbf{Output:} $P_0$ learns DNA profiles of $s_i \in S$ encrypted under $\mathcal{C}_H$.
            
            \begin{algorithmic}[1]
                \State $P_0$ generates private-key set $K' = \{k'_{s_i} : k_{s_i}$ and $k'_{s_i}$ are a key pair under $\mathcal{C}$ $\forall s_i \in S\}$.
                \State $P_0$ constructs a ZKP $\mathcal{Z}$ proving that it owns the private keys of exactly $m$ public keys in $K$. 
                \State $P_0$ transmits $K$ and $\mathcal{Z}$ to $P_1$ over a secure channel.
                \State $P_1$ verifies $\mathcal{Z}$. If verification fails, it sends back $\Phi$ to $P_0$ and the algorithm terminates. Else, it follows steps 6-12.
                \State $Y \gets \Phi$
                \For{$i \gets 1$ to $n$}
                    \State {$k \gets (K)_i$}
                    \State {$y_i \gets \mathcal{E}(\mathbf{d_i}^H, k)$} 
                    \State {$Y = Y \cup \{y_i\}$}
                \EndFor
                \State $P_1$ transmits $Y$ to $P_0$
                \State $\Psi_{\text{Suspects}} \gets \Phi$
                \For{$i \gets 1$ to $m$}
                    \State {$y \gets Y_{s_i}$,  $k' \gets K'_i$}
                    \State {$\mathbf{d_{s_i}}^H \gets \mathcal{D}(y, k')$} 
                    \State {$\Psi_{\text{Suspects}} = \Psi_{\text{Suspects}} \cup \{\mathbf{d_{s_i}}^H\}$}
                \EndFor
            \end{algorithmic}
        \end{algorithm}
    


\subsection{Encrypted DNA Profile of Crime Scene Sample}
Now that $\mathcal{I}$ has learnt the encrypted DNA profiles of the suspects, for the sake of DNA matching it must also learn the encryption of $\mathbf{d_s}$ under $\mathcal{C}_H$ from $\mathcal{L}$, without $\mathcal{L}$ learning anything about $\mathbf{d_s}$.

        \begin{algorithm}[H]
            \caption{Encrypt DNA Profile $\mathbf{\Pi}_{\textsc{EncryptDNA}}$($P_0, P_1$)}\label{alg:5}
            \textbf{Input:} $P_0$ holds the DNA profile of the sample obtained from the crime scene $\mathbf{d_s}$. 
                            $P_1$ holds the key $k_H$ to the homomorphic cryptosystem $\mathcal{C}_H = (\mathcal{E}_H, \mathcal{D}_H)$
                            used to encrypt DNA profiles.
                            
            \textbf{Output:} $P_0$ learns $\mathcal{E}_H(\mathbf{d_s})$.
            
            \begin{algorithmic}[1]                    
                \State{$P_0$ converts $\mathbf{d_s}$ into a vector $\mathbf{K} = T(\mathbf{d_s})$ and sets $\mathbf{K'} = \Phi$.}
                \For{$\kappa \in \mathbf{K}$}  
                    \State{$P_0$ samples a set $\Omega = \{r_1, r_2,\cdots, r_{\omega}\}$ ;
                    $\forall i, r_i \overset{\$}{\leftarrow} \mathbb{Z}^*_{p}$.
                    }
                    \State{$P_0$ samples $\gamma \overset{\$}{\leftarrow} \mathbb{I}^{\omega}_{\frac{\omega}{2}}$.}
                    \State{$\Gamma \gets \{r^{-1}_i : r_i \bullet r^{-1}_i \equiv 1$ (mod $p$) $ \forall r_i \in \Omega_{1..\gamma}\}$.}
                
                    \State{$\Omega \gets \{\kappa \bullet r_i$ (mod $p) : r_i \in \Omega\}$}
                    \For{$i \gets \omega - \gamma + 1$ to $\omega$}
                        \State{$r \overset{\$}{\leftarrow} \mathbb{Z}^*_{p}$}
                        \State{$\Gamma \gets \Gamma \cup \{r\}$}
                    \EndFor
        
                    \State{$P_0$ computes $\Omega_{\pi_1} = \pi_1(\Omega)$ using private permutation $\pi_1$.}
                    \State{$P_0$ computes $\Gamma_{\pi_2} = \pi_2(\Gamma)$ using private permutation $\pi_2$.}
                    \State{$P_0$ transmits $\Omega_{\pi_1}$ and $\Gamma_{\pi_2}$ to $P_1$.}
                    \State{$Y_{\Omega} \gets \Phi, Y_{\Gamma} \gets \Phi$}
                    \For{$i \gets 1$ to $\omega$} \begin{flushright}\Comment{Steps 15-21 take place at $P_1$}\end{flushright}
                        \State{$x_{\Omega} \gets (\Omega_{\pi_1})_i$}
                        \State{$x_{\Gamma} \gets (\Gamma_{\pi_2})_i$}
                        \State{$Y_{\Omega} \gets Y_{\Omega} \cup \{ \mathcal{E}_H(x_{\Omega}) \}$}
                        \State{$Y_{\Gamma} \gets Y_{\Gamma} \cup \{ \mathcal{E}_H(x_{\Gamma}) \}$}
                    \EndFor
                    
                    \State{$P_1$ computes $0^H = \mathcal{E}_H(0)$.}
    
                    \State{$P_1$ transmits $0^H$, $Y_{\Omega}$ and $Y_{\Gamma}$ to $P_0$.}
    
                    \State{$P_0$ computes $Y_{\Omega} = \pi^{-1}_1(Y_{\Omega})$ and $Y_{\Gamma} = \pi^{-1}_2(Y_{\Gamma})$.}
                    \State{$P_0$ selects random $t \overset{\$}{\leftarrow} \mathbb{I}^{\gamma}_1$.}
                    \State{$P_0$ computes $\mathcal{E}_H(\kappa) = (Y_{\Omega})_t \bullet (Y_{\Gamma})_t $.}
                    \State{$\mathbf{K'} \gets \mathbf{K'} \cup {\mathcal{E}_H(\kappa)}$}
                \EndFor
                \State{$P_0$ constructs $\mathbf{d_{s}}^H = \mathcal{E}_H(\mathbf{d_s}) = T^{-1}(\mathbf{K'}).$}
            \end{algorithmic}
        \end{algorithm}
 
  This condition is necessary because in the case when $\mathcal{L}$ learns $\mathbf{d_s}$, it can perform a DNA matching process locally between $\mathbf{D}$ and $\mathbf{d_s}$. In case $\mathcal{L}$ finds a match, it is possible that it can alert the perpetrator of the crime in exchange for some personal gain. To obtain the encryption of $\mathbf{d_s}$ without compromising its integrity, the protocol $\mathbf{\Pi}_{\textsc{EncryptDNA}}$ is executed with $P_0 = \mathcal{I}$ and $P_1 = \mathcal{L}$.

\subsection{Finding Similarities between DNA Profiles}
The algorithm $\mathbf{\Pi}_{\textsc{DNASimilarities}}$($P$) calculates the distance of the DNA profile of each suspect $s_i$ in $S$ encrypted under $\mathcal{C}_H = (\mathcal{E}_H, \mathcal{D}_H)$, with the encrypted DNA profile of the crime scene sample. $\mathbf{\Pi}_{\textsc{DNASimilarities}}$($P$) is called with $P = \mathcal{I}$.
        
        \begin{algorithm}[h]
            \caption{Find DNA Profile Similarities $\mathbf{\Pi}_{\textsc{DNASimilarities}}$($P$)}\label{alg:6}
            \textbf{Input:} $P$ holds an $m$-length vector of suspect DNA profiles $\Psi_{\text{suspects}}$ and the crime scene sample's DNA profile $\mathbf{d_s}$, each encrypted under a homomorphic cryptosystem $\mathcal{C}_H = (\mathcal{E}_H, \mathcal{D}_H)$. Each $\mathbf{d_{i}}^H \in \Psi_{\text{suspects}}$ and $\mathbf{d_{s}}^H$ is a $(t \times 2)$-dimension matrix.
                            
            \textbf{Output:} $P$ learns vector $\mathbf{\Delta}^H$ containing similarity scores of each $\mathbf{d_{i}}^H \in \Psi_{\text{suspects}}$ with $\mathbf{d_{s}}^H$, encrypted under $\mathcal{C}_H$.
            
            \begin{algorithmic}[1]
                \State {$\mathbf{\Delta}^H \gets \Phi$}
                \For{$i \gets 1$ to $m$}
                    \State{$\delta^H \gets 0^H$}
                    \State{$\mathbf{d_{s_i}}^H \gets (\Psi_{\text{suspects}})_i$}
                    \For{$j \gets 1$ to $t$}
                        \State{$\delta^H_1 \gets ([\mathbf{d_{s_i}}^H]_{j, 1} - [\mathbf{d_{s}}^H]_{j, 1}) \bullet ([\mathbf{d_{s_i}}^H]_{j, 1} - [\mathbf{d_{s}}^H]_{j, 1})$}
                        \vspace{1mm}
                        \State{$\delta^H_2 \gets ([\mathbf{d_{s_i}}^H]_{j, 2} - [\mathbf{d_{s}}^H]_{j, 2}) \bullet ([\mathbf{d_{s_i}}^H]_{j, 2} - [\mathbf{d_{s}}^H]_{j, 2})$} 
                        \vspace{1mm}
                        \State{$\delta^H \gets \delta^H + \delta^H_1 \bullet \delta^H_2 $}
                    \EndFor
                \EndFor
                \State{$\mathbf{\Delta}^H \gets \mathbf{\Delta}^H \cup \{\delta^H\}$}
            \end{algorithmic}
        \end{algorithm}

For each suspect’s DNA profile $\mathbf{d_{s_i}}^H$ encrypted under $\mathcal{C}_H$, its distance is calculated with the DNA profile of the crime scene sample $\mathbf{d_{s}}^H$ encrypted under $\mathcal{C}_H$. Note that here $0^H$ is the encryption of zero under $\mathcal{C}_H$.

\begin{figure*}[]
    \centering
    \includegraphics[width=0.8\textwidth]{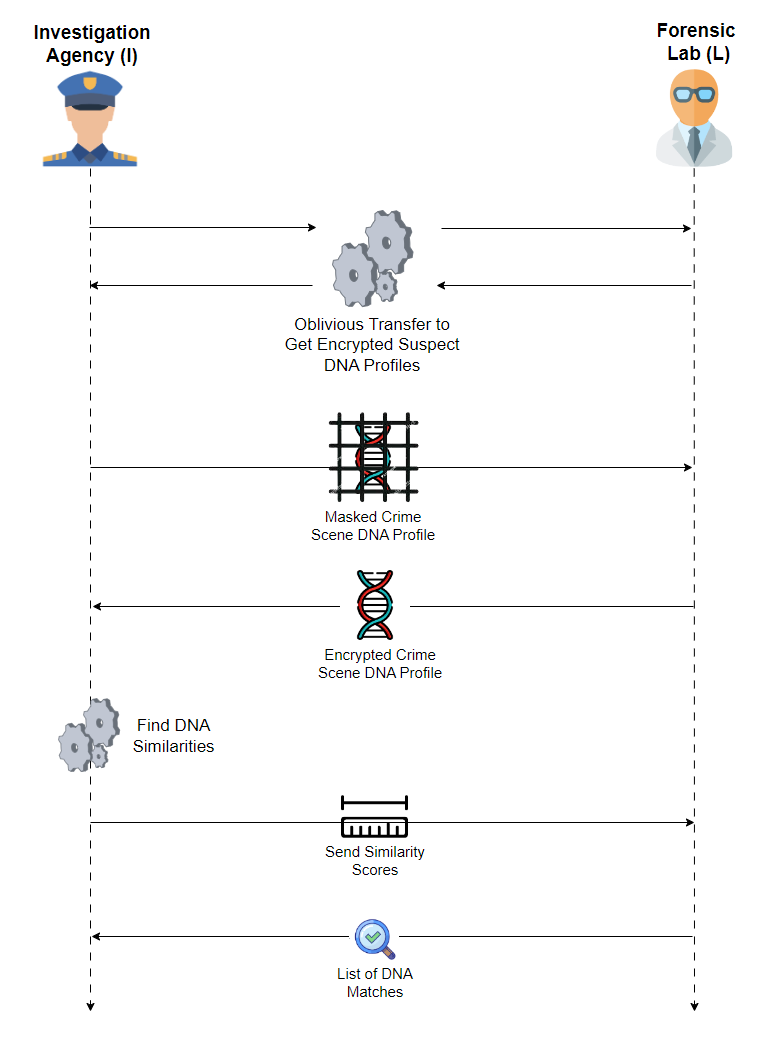}
    \caption{The PrivaMatch Scheme}
    \label{fig:real_scheme}
\end{figure*}

\subsection{Finding DNA Match}
In the final step, a match between the suspects' DNA profiles and the crime scene sample's DNA profile is found. The match is found by comparing the inverse-similarity scores calculated for the suspect DNA profiles in the previous step with a pre-defined match threshold $\tau$. This last step requires communication of the $\mathbf{\Delta}^H$ vector  to $L$ who can decrypt it and compare each $\delta_i$ with $\tau$. In this process, it is not revealed to $\mathcal{L}$ that which $\delta_i$ corresponds to which suspect. $\mathcal{L}$ only learns the values of $\delta_i$, from which it cannot learn anything about the DNA profiles $\mathbf{d_{s_i}}$ or $\mathbf{d_s}$ from which $\delta_i$ was calculated. $\mathcal{L}$ returns a list of indices $i$ to $\mathcal{I}$ such that $\delta_i$ lies within the match threshold $\tau$. Hence, $\mathcal{I}$ learns which suspects' DNA profiles match with the DNA profile of the sample obtained from the crime scene. Protocol $\mathbf{\Pi}_{\textsc{PrivaMatching}}$ is executed with $P_0 = \mathcal{I}$ and $P_1 = \mathcal{L}$.

        \begin{algorithm}[H]
            \caption{Find DNA Match $\mathbf{\Pi}_{\textsc{PrivaMatching}}$($P_0, P_1$)}\label{alg:7}
            \textbf{Input:} $P_0$ holds a vector $\mathbf{\Delta}^H$ of $m$ distances between the crime scene DNA profile $\mathbf{d_{s}}^H$ and each $\mathbf{d_{s_i}}^H \in \{\mathbf{d_{s_1}}^H, \mathbf{d_{s_2}}^H, \cdots, \mathbf{d_{s_m}}^H\}$, encrypted under the homomorphic cryptosystem $\mathcal{C}_H = (\mathcal{E}_H, \mathcal{D}_H)$. $P_0$ also holds a DNA match threshold $\tau$. $P_1$ holds the key $k_H$ of the cryptosystem $\mathcal{C}_H$.
                            
            \textbf{Output:} $P_0$ learns $\bigcup\{i\}$ $\forall$ $\mathbf{d_{s_i}}$ which are sufficiently close to $\mathbf{d_s}$, else $P_0$ learns $\Phi$.
            
            \begin{algorithmic}[1]
                \State{$P_0$ transmits $\mathbf{\Delta}^H$ and $\tau$ to $P_1$ over a secure channel.}
                \State{$\mu \gets \Phi$}
                \For{$i \gets 1$ to $m$}
                    \State {$\delta^H_i \gets (\mathbf{\Delta}^H)_i$}
                    \State {$\delta_i \gets  \mathcal{D}_H(\delta^H_i)$}
                    \State {$\delta \gets \sqrt[4]{\delta_i}$} 
                    \If{$\delta < \tau $}
                        \State{$\mu \gets \mu \cup \{i\}$}
                    \EndIf
                \EndFor
                \State $P_1$ transmits $\mu$ to $P_0$ over a secure channel.
                \State {$P_0$ learns the suspects $s_i$ for each $i \in \mu$.}
            \end{algorithmic}
        \end{algorithm}



\section{Security Analysis}
\label{sec:sec_anal}
We show that the proposed scheme, \textbf{PrivaMatch}, preserves privacy of suspects DNA profile against curious lab/investigator.
We assume that investigation agency $\mathcal{I}$ and forensic lab $\mathcal{L}$ are semi-honest parties.

\hspace{3cm}
\begin{theorem}
$\mathbf{\Pi}_{\textsc{getprofiles}}$ \textit{privately obtains the encrypted DNA profiles of the suspects}.
\end{theorem}

\textit{Proof:} The correctness of the protocol is realized as follows. $P_0$ supplies $n$ public-keys corresponding to each encrypted entity in the database $\mathbf{D}$. It is noted that for each $s_i \in S$, $P_0$ holds the private-key $k'_{s_i}$ corresponding to public-key $k_{s_i}$. Since $P_1$ encrypts every $\mathbf{d_{i}}^H \in \mathbf{D}$ with $k_i$, there is a one-to-one relation between all entities $\mathbf{d_{i}}^H \in \mathbf{D}$ and their corresponding encryption $\mathcal{E}(\mathbf{d_{i}}^H, k_i)$. As $P_0$ holds the private keys associated with each suspect, it computes $\mathbf{d_{s_i}}^H = \mathcal{D}(\mathcal{E}(\mathbf{d_{s_i}}^H, k_{s_i}), k'_{s_i})$ for all $s_i \in S$, where the integrity of each $\mathbf{d_{s_i}}^H$ is assured by the underlying cryptosystem $\mathcal{C}$. For security of the protocol, we require that $P_0$ does not learn the encrypted DNA profile of any entity $s'_i \notin S$, and $P_0$ should not learn the plaintext DNA profiles of any entity in $\mathbf{D}$. It is noted that by construction of $K'$ and the correctness of the ZKP $\mathcal{Z}$ that $P_0$ communicates to $P_1$, $\{ j : s_j \notin S \} \cap \{ j : k'_{s_j} \in K' \text{ is the private key for entity } s_j \} = \Phi$. Therefore, for each public-key $k_{s_j} \in K$ such that $s_j \notin S$, there exists no associated private-key $k'_{s_j}$ in $K'$. Therefore, $P_0$ does not learn $\mathbf{d_{s_j}}^H$ for any $s_j \notin S$ solely from $\mathcal{E}(\mathbf{d_{s_j}}^H, k_{s_j})$ due to the assumption that the underlying cryptosystem $\mathcal{C}$ used in the protocol is secure. We also require that $P_1$ does not learn which specific entities $P_0$ is suspecting. $P_1$ does not know for which specific $s_j$'s, $P_0$ holds private-keys $k'_{s_j}$ in $K'$ because of the zero-knowledge property of $\mathcal{Z}$. This concludes that the protocol $\mathbf{\Pi}_{\textsc{getprofiles}}$ privately obtains the encrypted DNA profiles of the suspects. \hfill$\square$ 

\vspace{0.5cm}
\begin{theorem}
\textit{The protocol} $\mathbf{\Pi}_{\textsc{EncryptDNA}}$ \textit{privately computes the encryption of a DNA profile $\mathbf{d_s}$ under $\mathcal{C}_H$.}
\end{theorem}

\textit{Proof:} We first prove the correctness of the protocol. We know that $\mathbf{d_s}$ is private to $P_0$ and the key $k_H$ of the homomorphic cryptosystem $\mathcal{C}_H$ is private to $P_1$. In the protocol, $P_0$ obtains the encryption $\kappa^H$ for each element $\kappa$ in $\mathbf{d_s}$ under $\mathcal{C}_H$.\\ 
$P_0$ samples $\omega$ random integers from the field $\mathbb{Z}^*_p$ to construct $\Omega$ and computes the multiplicative inverses of $\gamma (\gamma < \omega)$ of these integers to construct $\Gamma$. $P_0$ then modifies $\Omega$ by computing the modular product of the element $\kappa$ with each of the $\omega$ integers. By definition, for each $r_i \in \Omega$, $\kappa \bullet r_i \text{ (mod } p) \in \mathbb{Z}^*_p$. $P_0$ further randomizes $\Gamma$ by appending $\omega - \gamma$ random integers sampled from $\mathbb{Z}^*_p$ to $\Gamma$. Therefore, $|\Omega| = |\Gamma|$. Finally, $P_0$ randomizes the order of $\Omega$ and $\Gamma$ by applying random private permutations $\pi_1$ and $\pi_2$, respectively, to construct $\Omega_{\pi_1}$ and $\Gamma_{\pi_2}$. $P_0$ communicates $\Omega_{\pi_1}$ and $\Gamma_{\pi_2}$ to $P_1$. For each element $e$ in $\Omega_{\pi_1}$ and $\Gamma_{\pi_2}$, $P_1$ encrypts $e$ under to $\mathcal{C}_H$ to compute $e^H$. On receiving the encrypted $\Omega_{\pi_1}$ and $\Gamma_{\pi_2}$, $P_0$ applies the inverse transformations of $\pi_1$ and $\pi_2$ to restore the order of $\Omega$ and $\Gamma$, respectively. Now, $P_0$ selects a random index $t$ such that $r_t$ in $\kappa \bullet r_t \text{ (mod } p) \in \Omega$ and $q_t \in \Gamma$ are a multiplicative inverse pair under $\mathbb{Z}^*_p$. $P_0$ can select such a $t$ as it constructed $\Omega$ and $\Gamma$ at the beginning of the protocol. Using this index $t$, $P_0$ computes the product $(\Omega)_t \bullet (\Gamma)_t$ as follows:
\begin{flalign}
&(\Omega)_t \bullet (\Gamma)_t \\&= \mathcal{E}_H(\kappa \bullet r_t \text{ (mod } p)) \bullet \mathcal{E}_H(r^{-1}_t \text{ (mod } p))\\
                              &= \mathcal{E}_H(\kappa \bullet r_t \bullet r^{-1}_t \text{ (mod } p))\\
                              &= \mathcal{E}_H(\kappa \bullet 1 \text{ (mod } p))\\
                              &= \mathcal{E}_H(\kappa \text{ mod } p)\label{eq:kmodp}\\
                              &= \mathcal{E}_H(\kappa)\label{eq:k}
\end{flalign}

Equation \ref{eq:k} follows from \ref{eq:kmodp} due to the assumption in Section \ref{sec:intro} that all elements in a DNA profile are members of $\mathbb{Z}^*_p$. Therefore, $P_0$ learns the encryption of every $\kappa \in \mathbf{d_s}$ and constructs $\mathbf{d_{s}}^H$. We require that $P_1$ does not learn any $\kappa \in \mathbf{d_s}$ during any step of the protocol. Note that all $r_i \in \Omega$ were randomly chosen by $P_0$ from $\mathbb{Z}^*_p$. It follows that $\kappa \bullet r_i \text{ (mod }  p)$ is also randomly distributed in $\mathbb{Z}^*_p$ for all $i \in \mathbb{I}_1^{\omega}$. Therefore, $P_1$ cannot learn anything about $\kappa$ given this random distribution of $\kappa \bullet r_i \text{ (mod } p)$. Also, $P_1$ does not know the specific pairs of indices $(i, j)$ in $\Omega_{\pi_1}$ and $\Gamma_{\pi_2}$ such that the integers $r$ and $q$ used to calculate $(\Omega_{\pi_1})_i$ and $(\Gamma_{\pi_2})_j$ respectively form a multiplicative inverse pair under $\mathbb{Z}^*_p$. This is due to the fact that permutations $\pi_1$ and $\pi_2$ are random and private to $P_0$. Moreover, $P_1$ being a semi-honest adversary, it does not attempt a brute-force search to find the original order of $\Omega_{\pi_1}$ and $\Gamma_{\pi_2}$ to find these multiplicative inverse pairs. If $P_1$ could indeed find such a multiplicative inverse pair, it could have computed the value of $\kappa$ in the same way as $P_0$ did above by multiplying the respective elements from $\Omega_{\pi_1}$ and $\Gamma_{\pi_2}$. Therefore, $P_1$ does not learn anything about $\kappa$ from $\Omega_{\pi_1}$ or $\Gamma_{\pi_2}$. \text{    } \hfill $\square$\\

\begin{theorem}
\textit{The protocol} $\mathbf{\Pi}_{\textsc{PrivaMatchingReal}}$ \textit{securely finds a match between the encrypted crime scene DNA profile and the encrypted DNA profiles of the suspects.}
\end{theorem}

\textit{Proof:} For security, we require that $P_1$ does not learn anything about $d_s$ For each suspect $s_i \in S$, $P_1$ decrypts the distance-metric calculated by $P_0$ between $d_s$ and $d_{s_i}$ in the encrypted domain. This yields the following expression for the distance-metric $\delta$ for each suspect in plaintext:

    \begin{align}
    \label{eq:delta}    
        & \delta  = \sum_{\mathclap{j=1}}^{t} [([\mathbf{d_{s_i}} - \mathbf{d_s}]_{j, \eta}) \times ([\mathbf{d_{s_i}} - \mathbf{d_s}]_{j, \eta})\\ 
        & \hspace{1cm} \times ([\mathbf{d_{s_i}} - \mathbf{d_s}]_{j, \sigma}) \times ([\mathbf{d_{s_i}} - \mathbf{d_s}]_{j, \sigma})],
    \end{align}
\\

where $t$ is the number of slots in each DNA profile. Clearly, it is infeasible for $\mathcal{L}$ to determine $\mathbf{d_s}$ from $\delta$ in equation \ref{eq:delta} as it involves multiple modular additions and products, and due to the fact that $P_1$ does not know which specific $\mathbf{d_{s_i}}$ this metric has been calculated for. Therefore, $P_1$ can only check if $\delta$ lies within the match threshold $\tau$ and report all such matching indices in $\mathbf{\Delta}^H$ to $P_0$. Since, $P_0$ knows the relation between the indices of $\mathbf{\Delta}^H$ and corresponding suspect in $S$, it learns which specific suspects' DNA profiles match the DNA profile of the sample obtained from the crime scene. Therefore, the protocol correctly and securely finds the matches in DNA profiles. \hfill $\square$

\section{Performance Analysis}\label{sec:perf_anal}
We present analytical bounds on the run times for the different phases of the DNA matching process in Table \ref{tab:4}. In Table \ref{tab:4}, $n$ represents the size of the DNA database; $m$ represents the number of suspects; $t$ represents the number of slots in each DNA profile; $e$ and $d$ represent the average time to encrypt and decrypt an element in $\mathbb{Z}^*_{p}$ under a particular asymmetric-key cryptosystem, respectively; $h_a$ and $h_m$, respectively, represent the average time to add and multiply two elements encrypted under a homomorphic cryptosystem; $s$ represents the average time for sampling a random element from $\mathbb{Z}^*_{p}$; and $i$ represents the average time to calculate the inverse of an element in $\mathbb{Z}^*_{p}$. 

\begin{table}[t]
    \centering
    \caption{Computation Complexities of the proposed scheme}
    \vspace{1mm}
    \begin{tabular}{|c|c|}
        \hline
        \textbf{Phase} & \textbf{Computational Complexity}              \\ \hline
        GetDNAProfiles                   & $\Theta(gn + etn + dmt)$           \\ \hline
        EncryptDNA                       & $\Theta(t\{(s\omega + i\gamma + s(\omega - \gamma) + e\omega + h_m\})$           \\ \hline
        DNASimilarities                   & $\Theta(tm\{2(2h_s + h_m) + h_m + h_s\})$           \\ \hline
        PrivaMatching                   & $\Theta(md)$           \\ \hline        
    \end{tabular}
    \label{tab:4}
\end{table}

\begin{table*}[t]
\centering
\caption{Running times for different functionalities in the PrivaMatch protocol}
\begin{tabular}{|cc|cccccc|}
\hline
\multicolumn{2}{|l|}{\textbf{\begin{tabular}[c]{@{}l@{}}DNA Database and\\  Profile Parameters\end{tabular}}}                                                       & \multicolumn{6}{c|}{\textbf{Time to Execute Functionality (in milliseconds)}}                                                                        \\ \hline
\multicolumn{1}{|c|}{\textbf{\begin{tabular}[c]{@{}c@{}}Profile \\ Slots (t)\end{tabular}}} & \textbf{\begin{tabular}[c]{@{}c@{}}Database\\  Size (n)\end{tabular}} & \multicolumn{1}{c|}{\textbf{\begin{tabular}[c]{@{}c@{}}Oblivious \\ Transfer \\ (OT) Key \\ Generation\end{tabular}}} & \multicolumn{1}{c|}{\textbf{\begin{tabular}[c]{@{}c@{}}OT DNA\\  Profile\\ Encryption\end{tabular}}} & \multicolumn{1}{c|}{\textbf{\begin{tabular}[c]{@{}c@{}}OT DNA\\  Profile\\ Extraction\end{tabular}}} & \multicolumn{1}{c|}{\textbf{\begin{tabular}[c]{@{}c@{}}Crime Scene \\ DNA\\ Profile \\ Encryption\end{tabular}}} & \multicolumn{1}{c|}{\textbf{\begin{tabular}[c]{@{}c@{}}Find DNA \\ Similarities\end{tabular}}} & \textbf{\begin{tabular}[c]{@{}c@{}}Find DNA \\ Match\end{tabular}} \\ \hline
\multicolumn{1}{|c|}{10}                                                                    & 10                                                                    & \multicolumn{1}{c|}{18552}                                                                                            & \multicolumn{1}{c|}{254}                                                                             & \multicolumn{1}{c|}{358}                                                                             & \multicolumn{1}{c|}{3951}                                                                                        & \multicolumn{1}{c|}{555}                                                                       & 13                                                                 \\ \hline
\multicolumn{1}{|c|}{50}                                                                    & 10                                                                    & \multicolumn{1}{c|}{21890}                                                                                            & \multicolumn{1}{c|}{1461}                                                                            & \multicolumn{1}{c|}{1884}                                                                            & \multicolumn{1}{c|}{17648}                                                                                       & \multicolumn{1}{c|}{2804}                                                                      & 12                                                                 \\ \hline
\multicolumn{1}{|c|}{100}                                                                   & 10                                                                    & \multicolumn{1}{c|}{20184}                                                                                            & \multicolumn{1}{c|}{2879}                                                                            & \multicolumn{1}{c|}{3765}                                                                            & \multicolumn{1}{c|}{30750}                                                                                       & \multicolumn{1}{c|}{5455}                                                                      & 12                                                                 \\ \hline
\multicolumn{1}{|c|}{10}                                                                    & 50                                                                    & \multicolumn{1}{c|}{109889}                                                                                           & \multicolumn{1}{c|}{1423}                                                                            & \multicolumn{1}{c|}{375}                                                                             & \multicolumn{1}{c|}{4127}                                                                                        & \multicolumn{1}{c|}{560}                                                                       & 13                                                                 \\ \hline
\multicolumn{1}{|c|}{10}                                                                    & 100                                                                   & \multicolumn{1}{c|}{263612}                                                                                           & \multicolumn{1}{c|}{2921}                                                                            & \multicolumn{1}{c|}{400}                                                                             & \multicolumn{1}{c|}{4435}                                                                                        & \multicolumn{1}{c|}{626}                                                                       & 13                                                                 \\ \hline
\end{tabular}
\label{tab:5}
\end{table*}

We use a set of randomly generated DNA profiles of the form described in equation \ref{eq:1} to populate the profile database $\mathbf{D}$ owned by $\mathcal{L}$. We also simulate a DNA profile obtained from the crime scene $\mathbf{d_s}$ and set its ownership to $\mathcal{I}$. Using different sets of people suspected by $\mathcal{I}$, we simulate the \textbf{PrivaMatch} scheme in the ideal and real world settings. In the implementation of the scheme, we have used a Galois field $\mathbb{GF}(p)$ of order $p = 2^{16} + 1$ for all arithmetic operations. The Galois field is simulated using the \verb|galois| Python library \cite{Hostetter_Galois_2020} and all homomorphic encryption operations are simulated using the \verb|PyFhel| library \cite{ibarrondo2021pyfhel} implementing the BGV cryptosystem. For the oblivious transfer scheme described in Section \ref{sec:scheme}, we have used the RSA cryptosystem implemented using the \verb|rsa| Python library \cite{rsa}. The simulations have been carried out on a system with an Intel Xeon CPU core with 2 virtual CPUs and a total RAM size of 12.7 GB. The running times for the complete DNA extraction and matching process (in the real-world model) as explained in the previous sections has been presented in Table \ref{tab:5}. The code for the implementation can be found in the repository at \texttt{\href{https://github.com/sankha555/privamatch}{github.com/sankha555/privamatch}}.

\section{Conclusions}\label{sec:concl}
We proposed a scheme named PrivaMatch for private DNA matching in forensic investigations where the DNA profile of a sample obtained from a crime scene is matched against the DNA profiles of a list of suspects. We have addressed a number of concerns related to privacy and integrity of the investigation process through this scheme operating in a multi-party adversarial setting. The privacy requirement that the owner of the DNA profile database does not learn about the specific suspects being targeted by the investigation agency is satisfied by the completeness of the oblivious transfer scheme described in the first phase of the scheme. Moreover, the privacy of the suspects' DNA profiles is maintained by the fact that all the DNA profiles communicated to the investigation agency are encrypted under a homomorphic encryption. In the second phase of the scheme, the DNA profile of the crime scene sample is privately encrypted under the same homomorphic cryptosystem without the owner of the cryptosystem's key (the forensic laboratory) learning anything about the sample's DNA profile. This encrypted DNA profile is compared against each of the encrypted DNA profiles of the suspects procured in the first phase to calculate a distance-metric. The distance-metric is used to decide if the particular suspect's DNA profile is ``sufficiently'' similar to the crime scene sample's DNA profile, which is the intended objective of the DNA matching scheme. In the entire process, neither of the parties (the investigation agency and the forensic lab) learns about the private data of the other parties (the crime scene DNA profile and the suspects' DNA profiles, respectively), therefore, satisfying the requirements of a sound and private secure multiparty computation protocol. As a result, the 
privacy and identity of the individuals registered in the DNA profile database is not compromised. In addition, sensitive information crucial to the smooth operation of the investigation process is not leaked to external parties.



\section{Funding and Competing Interests}
Funding: The authors declare that the project was not funded by any organization or individual.
Conflict of Interest: The authors declare that they have no conflict of interest.

\section{Data Availability}
Data sharing not applicable to this article as no datasets were generated or analyzed during the current study




\end{document}